\documentclass[aps,numerical,graphicx,reprint,pre]{revtex4-1}
\usepackage{amsmath}
\usepackage{graphicx}
\draft 

\begin{document}

\def\Journal#1#2#3#4{{#1 }{\bf #2, }{ #3 }{ (#4)}}

\def\BiJ{ Biophys. J.}
\def\Bios{ Biosensors and Bioelectronics}
\def\LNC{ Lett. Nuovo Cimento}
\def\JCP{ J. Chem. Phys.}
\def\JAP{ J. Appl. Phys.}
\def\JMB{ J. Mol. Biol.}
\def\JPC{ J. Phys: Condens. Matter}
\def\CMP{ Comm. Math. Phys.}
\def\LMP{ Lett. Math. Phys.}
\def\NLE{{ Nature Lett.}}
\def\NPB{{ Nucl. Phys.} B}
\def\PLA{{ Phys. Lett.}  A}
\def\PLB{{ Phys. Lett.}  B}
\def\PNAS{Proc. Natl. Am. Soc.}
\def\PRL{ Phys. Rev. Lett.}
\def\PRA{{ Phys. Rev.} A}
\def\PRE{{ Phys. Rev.} E}
\def\PRB{{ Phys. Rev.} B}
\def\PD{{ Physica} D}
\def\ZPC{{ Z. Phys.} C}
\def\RMP{ Rev. Mod. Phys.}
\def\EPJD{{ Eur. Phys. J.} D}
\def\SAB{ Sens. Act. B}
\title{
Why zero-point quantum noise cannot be detected at thermal equilibrium:
Casimir force and zero-point contribution in the fluctuation-dissipation theorem  
}
\author{Lino Reggiani}
\email{lino.reggiani@unisalento.it}
\affiliation{Dipartimento di Matematica e Fisica, ``Ennio de Giorgi'',
Universit\`a del Salento, via Monteroni, I-73100 Lecce, Italy}
\affiliation{CNISM,  Via della Vasca Navale, 84 - 00146 Roma, Italy}

\author{Eleonora Alfinito}
\affiliation{Dipartimento di Ingegneria dell' Innovazione, Universit\`a del
Salento, via Monteroni, I-73100 Lecce, Italy} 
\affiliation{CNISM,  Via della
Vasca Navale, 84 - 00146 Roma, Italy}
\date{\today}
\begin{abstract}
The role played by zero-point contribution, also called quantum noise or vacuum fluctuations, in the quantum expression of the fluctuation-dissipation theorem (FDT) is a long-standing open problem widely discussed by the physicist community  since its announcement by Callen and Welton pioneer paper 
of 1951 \cite{callen51}.  
From one hand, it has the drawbacks of: (i) the expectation value of its energy is infinite, (ii) it produces an ultraviolet catastrophe of the noise power spectral density and, (iii) it lacks of an experimental validation under thermal equilibrium conditions. 
From another hand, by imposing appropriate boundary conditions and eliminating divergences by regulation techniques,  vacuum fluctuations are the source of an attractive force between opposite conducting  plates, firstly predicted by Casimir in 1948 \cite{casimir48} and later validated experimentally with increasing accuracy.
As a consequence, a quantum formulation of FDT should account for the presence of the Casimir force and of its consequences.
In this letter we show that  at thermal equilibrium the Casimir force should be balanced by the mechanical reaction of the physical system.
As a consequence,  no zero-point spectrum can be detected and the power spectrum emitted by the physical system is the same of that calculated by Planck in 1901 
\cite{planck01} for a black-body.
Accordingly, the experimental validation of the standard expression of the quantum FDT \cite{callen51} is prevented in favor of the Nyquist expression that includes the Planck factor \cite{nyquist28}. 
\end{abstract}
\maketitle 
%
The formulation of the quantum fluctuation dissipation theorem (FDT) traces back to the theory of the black body radiation spectrum given by Planck in 1901 and 1912 \cite{planck01, planck12}. Then, the theorem was extended to the case of electrical fluctuations in conductors by Nyquist in 1928 \cite{nyquist28} and generalized by Callen and Welton in 1951 \cite{callen51} on the basis of a rigorous quantum mechanics approach. The theorem was further elaborated by Kubo in 1957 \cite{kubo57} using a linear-response and correlation-function approach. To date the quantum FDT can be formulated on the basis of quantum statistics using the Gibbs ensemble method. 
This method enables to present a unified theory of the quantum FDT that relates the energy exchange between a  thermal reservoir at temperature $T$ and a physical system (e.g. a black-body cavity or a given  volume filled with a conductor) weakly coupled each other as described by a canonical or a  grand-canonical ensemble at thermodynamic equilibrium \cite{reggiani16}.
The theorem can be formulated also within a microcanonical ensemble \cite{bonanca08}, but, in this case, being absent any energy exchange,  it establishes at a kinetic level a relation between linear-response function and correlation functions in terms of the width of the total energy of the ensemble instead of the temperature.  
\par
Common to all quantum formulations, the power spectral-density of the radiation emitted by the physical system includes the contribution of the energy corresponding to the ground state of the harmonic oscillator, $hf/2$, with $h$ the Planck constant and $f$ the photon frequency, the so called zero-point energy contribution or, depending on the context, quantum noise \cite{gardiner91,clerk10} or vacuum fluctuations \cite{beck07}.
The physical justification and the detection or less of zero-point energy were and still are a source of controversial discussions among the physicists community \cite{klimontovich87,ginzburg87,tatarskii87,senitzky93,lesovik97,levinson00,levinson03,reggiani12,kish15}, that represents an outstanding unsolved problem.
\par
In this letter we propose a solution of this open problem.
To this purpose, we recall that
the zero point contribution to thermal noise is related to the finite size or boundary conditions of the physical system \cite{kardar99,schmidt08},
which, in turn, is at the origin of the Casimir force acting on the structure.
of the physical system. 
\par
From theory, the presence of zero-point energy in thermal noise originates from the fundamental uncertainty principles, that imply non-commuting properties at different times of the quantum operator associated with the given observable. 
As a consequence, zero-point contribution  should be considered unavoidable, despite of existing criticism \cite{klimontovich87, ginzburg87, tatarskii87, senitzky93,lesovik97,clerk10,levinson00,levinson03,reggiani12,kish15}.
\par
From experiments, to our knowledge this contribution has not been  validated by  noise measurements at thermodynamic equilibrium, while out of equilibrium experimental evidence was claimed for the case of a resistivity shunted Josephson junction \cite{koch80}, and noise squeezing \emph{ via} a Josephson-parametric amplifier \cite{movshovich90}.
Rather, some researchers  suggested the impossibility of obtaining such a direct evidence by conjecturing that zero-point energy does not represent an exchangeable energy \cite{davytov86,levinson00}.
The celebrated Planck 1901  formula \cite{planck01} for the interpretation of black-body radiation spectrum, where zero-point energy is omitted, is a crucial example in support of the above conjecture.
Indeed, the excellent agreement between theory and experiments is obtained without including the zero-point energy.
\par
On the other hand, the existence of the zero-point energy received theoretical and experimental evidence within the Casimir effect, a genuine 
quantum phenomenon announced in 1948 \cite{casimir48} and predicting the attraction between  conducting parallel thin-plates of given surface area $A$ placed at a distance $L$ in free space.
Due to the exclusion of electromagnetic modes between the plates as compared to free space, the Casimir force, $F_C$, associated with the expected value of the zero-point electromagnetic  energy, $U_{zp}$, defined as 
\begin{equation}
U_{zp} = \frac{1}{2} \sum hf
\end{equation}
where the sum is extended over all the photon modes, writes \cite{casimir48}
\begin{equation}
F_C
=- \frac{\pi h c A}  {480 L^4} 
\end{equation}
with  $c$ the light speed in vacuum.
\par
The Casimir effect,  received experimental validation with increasing precision  \cite{lamoreaux97,mohideen98,decca03}, and was further elaborated by Liftshitz \cite{liftshitz56} to predict also a repulsion effect for more complicate systems consisting of dielectrics and more realistic metal plates 
\cite{imry05,rodriguez11}. 
(For an updated bibliography on the Casimir effect the reader can refer to Ref. \cite{babb}.)
\par 
The essence of FDT can be expressed by considering the thermal noise-power per unit  bandwidth, $P(f)/\Delta f$, radiated into a single mode of the electromagnetic field by a conductor of volume $A \dot L$ (taken as the physical system) in contact with a thermal reservoir at temperature $T$.
The thermal noise-power is defined by the ratio between the variance of voltage (current) fluctuations and the real part of the conductor impedance (admittance). 
Here, for simplicity of calculations, the conductor is taken as a particular case of an ideal black-body that is filled by a homogeneous conducting material, with length $L$ in the $z$-direction, cross section $A$, and with the length much smaller than the transverse directions. 
We notice that for an ideal black-body the  impedance coincides with the vacuum resistance given by $R_{vacuum}=\mu_0 c$, with $\mu_0$ the vacuum permeability.
In particular, the vacuum resistance is related to the Von Klitzing fundamental unit of resistance \cite{klitzing80}, $h/e^2$, by $R_{vacuum}=2 \alpha h/e^2$ with $\alpha$ the fine structure constant. 
Accordingly, the FDT formulated by Callen and Welton recovers the Planck 1912 \cite{planck12} second formulation of the black-body radiation spectrum which includes zero-point contribution. 
\par 
We notice, that at the level of Planck 1901  first formulation of black-body radiation  spectrum \cite{planck01} and of Nyquist 1928 theorem \cite{nyquist28} on the electrical fluctuations in conductors, the vacuum fluctuation contribution to the power spectrum was missed.
The quantum mechanical approach of Callen and Welton in 1951 \cite{callen51}, as well as that of Kubo in 1957 \cite{kubo57}, generalized the Nyquist theorem by including the zero-point contribution on the basis of a rigorous use of quantum mechanics. However, their formulations were not extended explicitly to the case of the black-body radiation. Furthermore, Casimir force and its implications were not considered.
The above results are conveniently summarized by the following relations:
\begin{equation}
\frac{P_{Planck}(f)}{\Delta f} =
K_BT \frac{x}{e^x-1}
\end{equation}
\begin{equation}
\frac{P_{Nyquist}(f)}{\Delta f} =
K_BT 
\end{equation}
\begin{equation}
\frac{P_{CWK}(f)}{\Delta f} =
K_BT  \frac{x}{2} coth(\frac{x}{2})=
\frac{P_{Planck}(f)}{\Delta f} + \frac{hf}{2} 
\end{equation}
%
with $P_{CWK}(f)$ referring to Callen-Welton \cite{callen51} and Kubo 
\cite{kubo57} formulation,   $x=hf/K_BT$ and $K_B$ the Boltzmann constant.
\par
We stress that the split into two contribution of $P_{CWK}(f)$ as given by the last expression in the r.h.s. of Eq. (5) is of most physical importance.
Indeed, the first-Planck contribution represents a property of the coupling between the thermal reservoir and the physical system at thermal equilibrium as expressed by the detailed energy-balance related with the microscopic process of energy absorption (emission) between the thermal reservoir and the physical system. 
As such, it is a universal function of the temperature which takes a finite value at any frequency.
Accordingly, it vanishes at $T=0$,  it is independent of the external shape of the  physical system and of the conducting material inside to the physical system,
its spectrum can be directly measured by standard experimental techniques in a wide range of frequencies, typically from mHz to THz, and excellent agreement between theory and experiments is a standard achievement.
\par
By contrast, the second zero-point energy contribution is a property of the vacuum and from its definition in Eq. (1): its expectation value diverges, it does not vanish at $T=0$,  it has never been measured directly  but only through its effects as evidenced by attractive or repulsive forces acting between finite parts of the physical system, as predicted for the simple case of two parallel conducting plates by Casimir \cite{casimir48} and reported in Eq. (2).
\par
The essential difference between the above two contributions is better explained when considering the total energy obtained by summation over all the photon modes of each contribution.
For the Planck term, the summation is easily performed and gives the well-known result:
\begin{equation}
U_{Planck} = \sum K_BT \ \frac{x}{e^x-1} = 3 \frac{\zeta(4)}{\zeta/3)} N K_B T
\end{equation}
with $\zeta(n)$ the Riemann $\zeta$ function and $N$ the expected number of phonon inside the volume of the given system.
\par
For the zero-point term the summation gives the expectation value of the energy of the electromagnetic field as defined in Eq. (1). Written in this way it diverges and to obtain a finite value it is necessary  to make use of renormalization or regularization techniques to avoid divergences and to include boundaries conditions related: (i)  to the shape of the physical system, (ii) to the material inside the physical system.
Calculations are not easy to be performed \cite{ederly06,schmidt08,auletta09}, and here we report the simple but significant  case considered by Casimir \cite{casimir48} and further confirmed by more detailed mathematical approaches \cite{milton}: 
\begin{equation}
U_{C} = - \frac{\pi A h c }{1440 L^3} 
\end{equation}
The negative value of the Casimir energy $U_{C}$ corresponds to an attractive force (the Casimir force) between opposite conducting plates $F_C=- dU_{C}/dL$ as reported in Eq. (2). 
As a consequence of this energy, the physical system is not mechanically stable and thus tends to implode \cite{lebowitz69}.
\par
Here, we propose that the inclusion of vacuum fluctuations should be combined with the presence of the Casimir force acting on the structure defining the physical system of interest, and of its consequences as detailed in the following. 
The Casimir effect, by converting vacuum fluctuations into an attractive force between opposite terminal-surfaces (contacts) implies a stress on the  structure of the physical system.
In turn,  this stress  is stored into an energy strain $U_s$ by the body structure.
To keep the stability of the physical system the condition $U_s = - U_{C}$ is a sufficient one.
Accordingly, by accounting for the Casimir force and the strain energy stored by the system, at thermodynamic equilibrium the zero-point energy contribution is washed out.
Thus, the original Planck 1901 \cite{planck01} expression for the black-body radiation emission as well as the Nyquist relation replacing $K_BT$  by the Planck factor are recovered, in full agreement with experimental evidence as originally suggested by Nyquist himself \cite{nyquist28}. .
\par
In conclusion, we propose that in a quantum derivation of the  FDT the energy contribution of vacuum fluctuations should be balanced at a macroscopic level by the associated strain energy stored by the physical system under test.
Hence, vacuum fluctuations do not contribute to the determination of 
the noise power-spectrum  emitted by the physical system.
Accordingly, as reported by Eq. (3), Planck law of 1901 \cite{planck01} for black-body radiation spectrum can be interpreted as the quantum FDT  \cite{clerk10} with: (i) the black-body radiator playing the role of the vacuum impedance,  (ii) the bandwidth, $\Delta f = c/L$, being the inverse of the ballistic transit time of a photon travelling between contacts and, (iii) the external boundaries playing the role of a thermal reservoir.
The same Eq. (3) recovers the quantum Nyquist formula for conductors by replacing the vacuum admittance (or impedance) with that of a real conductor that includes capacitance and kinetic-induction effects as well as the natural bandwidths associated with the carrier collision-time or the dielectric relaxation-time according to constant voltage or constant-current measurement conditions, as detailed in \cite{reggiani16}. 
\par
We finally remark, that the above results can be summarized by defining a total energy of the phonon gas inside the physical system, $U_T$, as a sum of three contributions
\begin{equation}
U_T= U_{Planck} + U_{C} + U_s 
\end{equation}
where 
\begin{equation}
 U_s = - U_{C}  
\end{equation}
is the strain energy stored  by  the  physical system balancing  the  stress energy due to  the Casimir force and ensuring
the stability of the physical system.
We notice that Eq. (8) is reminiscent of the van der Waals corrections to the ideal classical gas.
\par
In summary, we confirm that quantum noise cannot be detected  in the form predicted by the quantum FDT \cite{planck12,callen51,kubo57} since at thermal equilibrium the associated Casimir force is exactly cancelled by the mechanical reaction of the physical system.
\begin{acknowledgments} 
{Dr. T. Kuhn from M\"unster University, Germany, is thanked for valuable discussions on the subject.}
\end{acknowledgments} 

\end{document}